\documentclass{article}
\usepackage{graphicx} 
\usepackage{geometry}
\usepackage{cite}
\usepackage[usenames]{color}
\usepackage{colortbl}
\usepackage{amsmath}
\usepackage{subfigure}
\usepackage{cleveref}
\usepackage{authblk}
\usepackage[shortlabels]{enumitem}

\title{Observation of droplets in dimer model on a triangular lattice}
\author[1]{Daria Shvalyuk}
\author[1,2]{Anton Nazarov}
\affil[1]{Saint Petersburg State University, 7/9 Universitetskaya nab., 199034 Saint Petersburg, Russia}
\affil[2]{Beijing Institute of Mathematical Sciences and Applications (BIMSA),
Bejing 101408, People’s Republic of China}
\date{}                     
\begin{document}

\maketitle

\begin{abstract}
In this paper, we consider the formation of droplets in the dimer model on a triangular lattice. The droplets in the dimer model are superposition polygons formed as two overlapping  configurations of dimers: constant and movable. We demonstrate that specific local energies of dimers and low temperatures lead to the emergence of a macroscopic droplet. The motivation for this study was the phenomenon of the formation of equilibrium droplets under certain conditions in the Ising model within the framework of the Dobrushin-Kotecký-Shlosman theory. Due to the deep connections between the Ising model and the dimer model, similar behaviour was expected.

\begin{description}
\item[Key words:]
dimer model, Ising model, Dobrushin-Kotecký-Shlosman theory, Monte Carlo simulation.
\end{description}
\end{abstract}

\section{Introduction}

The dimer model was developed to describe the adsorption of diatomic molecules on crystal surfaces~\cite{Roberts-1935}. Further, Fowler and Rushbrooke considered the problem of the equilibrium state of monomer-dimer systems as a model of liquid mixtures in which the molecules are unequal in size~\cite{Fowler-Rushbrooke-1937}. Henceforth the dimer model has found a wide range of applications in physics and mathematics~\cite{Kasteleyn-1963, Kenyon-Okounkov-2006, Kenyon-Okounkov-Sheffield-2003, Cimasoni-Reshetikhin-2007, Cimasoni-Reshetikhin-2008}.
 
A dimer configuration is a perfect matching in a graph. A wide class of various statistical models is reduced to dimer model~\cite{Fan-Wu-1970}. For example, due to the equivalency of the Ising model to the dimer model on the special lattice, the problem of calculation of partition function and correlation functions can be reformulated in terms of dimers~\cite{Kasteleyn-1963, Fisher-1966, Temperley-Fisher-1961, McCoy-Wu-1973}.

The phenomenon of droplet formation in the two-dimensional Ising ferromagnet within the Dobrushin-Kotecký-Shlosman (DKS) theory~\cite{Dobrushin-Kotecky-Shlosman-1992} is widely known and this result also can be generalised to the Ising model in higher dimensions\cite{Dobrushin-1973, Fröhlich-Pfister-1987, Ioffe-Ott-Shlosman-Velenik-2022, Pfister-Velenik-1997}. A droplet in the Ising model is a cluster of pluses or minuses (it depends on the boundary conditions) immersed in the opposite phase and separated from it by a certain curve. Originally, Wulff~\cite{Wulff-1901} formulated the problem of deriving the equilibrium shape of the crystal, which can be reduced to solving a variational problem (the Wulff construction) assuming surface tension to be known. To introduce interface in the Ising model, it is necessary to set up boundary conditions: pluses are in the upper part of the boundary, and minuses are in the lower part of the boundary. These conditions lead to the Ising model's configuration with an interface separating pluses and minuses. Surface tension is defined as the difference between free energies of two Ising configurations: the first configuration has all pluses on the boundary and the second configuration has pluses and minuses on the boundary providing interface~\cite{Abraham-Gallavotti-Martin-Lof-1971, Abraham-Reed-1977}. The equilibrium shape of the crystal corresponds to minimising the surface tension. As shown in~\cite{Dobrushin-Kotecky-Shlosman-1992}, if the lattice is large enough, the contour surrounding the droplet is described by the Wulff curve~\cite{Herring-1951} with high accuracy. According to DKS theory, phase separation occurs only when the temperature is sufficiently low. A range of works was devoted to the proof that phase coexistence is possible at any temperature below the Curie point~\cite{Ioffe-Schonmann-1998, Ioffe-1995, Greenberg-Ioffe-2005, Campanino-Ioffe-Velenik-1995, Ioffe-1994}.
Additionally, similar processes are observed, for example, in the Potts model and Blume-Capel model~\cite{Bricmont-Lebowitz-1987, Coninck-Messager-Miracle-Sole-Ruiz-1988, Hryniv-Kotecky-2002, Messager-Miracle-Sole-Ruiz-Shlosman-1991}.

In this work, we study a droplet formation in the dimer model. Droplets in the dimer model are superposition polygons from the overlapping reference dimer configuration and optimal dimer configuration at a given temperature.

It is worth noting that the behaviour of the system is determined by the lattice type (bipartite or non-bipartite). In the case of a bipartite graph, one can colour "black" and "white" vertices where the colour of the adjacent vertices of each vertex is different from the colour of itself. For such graphs, it is possible to introduce the height function~\cite{Gartner-2014}. As it was shown in~\cite{Kenyon-Okounkov-Sheffield-2003}, the dimer model on the bipartite lattice with periodic weights of edges has three phases: liquid, which is characterized by quadratic decay of correlators and fluctuations of height function logarithmically depending on the distance between points, frozen where fluctuations are absent and value of height function is deterministic for arbitrary distances between points, and gaseous with limited fluctuations and exponential decay of correlators.

We study the phase separation in the dimer model at low temperatures on the triangular lattice. Non-bipartiteness of the lattice leads to the inability to implement the standard height function. Instead, the height function modulo two is constructed here. Because of the principle of equivalence of ensembles, the dynamics of dimers can be formulated in terms of a grand canonical ensemble introducing the external field or (small) canonical ensemble. Here, the latter was selected. The dynamics was constructed in such a fashion that flips of the dimers did not change the area of the polygon within the contour. It is an analogue of Kawasaki sampling for Ising model~\cite{Landau-Binder-2000}. Previously, in the work~\cite{Fendley-Moessner-Sondhi-2002}, the dimer model on the triangular lattice with three types of edges was studied. Authors considered horizontal and two leaned edges, however, they deformed the triangular lattice so that triangles were rectangular and, as a result, one of the leaned dimers was strictly vertical. It was established that this system undergoes the phase transition when it reduces to the model in the square lattice when one of the weights of edges is equal to zero. Here, we introduce four types of edges (horizontal coinciding with reference configuration, horizontal non-coinciding with reference configuration, right-leaned, and left-leaned), whose energies were selected in such a way as to form a macroscopic droplet.

\section{Theory and method}
\subsection{Model definition}

Model is studied as a planar, simple, and non-bipartite graph $\Gamma$ on a triangular lattice with free boundary conditions. A \textit{dimer configuration} is a perfect matching in the graph $\Gamma$ which is a subset $\mathcal{D}$ of edges of the $\Gamma$ such that every vertex of the graph relates to exactly one edge in $\mathcal{D}$. Note, that the implementation of the concept of perfect matching is allowed only for graphs with an even number of vertices.

Consider \textit{energy} of a dimer configuration $\mathcal{D}$ in the absence of external field:
\begin{equation}
    E(\mathcal{D}) = \sum_{e \in \mathcal D} E(e),
\end{equation}
here $E(e)$ is energy of a dimer.

\textit{Boltzmann distribution} for dimer configurations is expressed as follows:
\begin{equation}
    \begin{aligned}
        \text{Prob}(\mathcal{D}) = \frac{1}{Z}\prod_{e \in \mathcal D} w(e), \\
        Z = \sum_{\mathcal D}\prod_{e \in \mathcal D} w(e), \\
        w(e) = \exp \bigg(-\frac{E(e)}{k_\text{B}T} \bigg),
    \end{aligned}
\end{equation}
here $Z$ -- the partition function, $w(e)$ -- Boltzmann weights of edges, $k_\text{B}$ -- the Boltzmann constant ($k_\text{B}$ is set to 1), $T$ -- temperature.

\begin{figure}[ht!]  
    \vspace{2ex}
    \centering
    \subfigure[]{
    \includegraphics[width=0.4\linewidth]{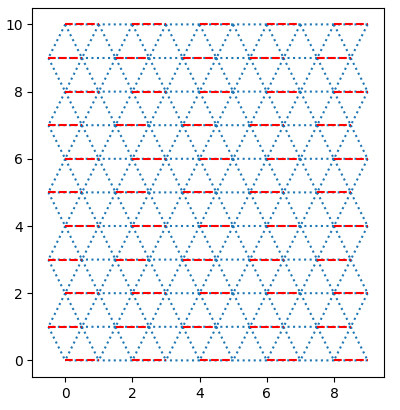} 
    \label{fig:reference} } 
    \hspace{4ex}
    \subfigure[]{
    \includegraphics[width=0.4\linewidth]{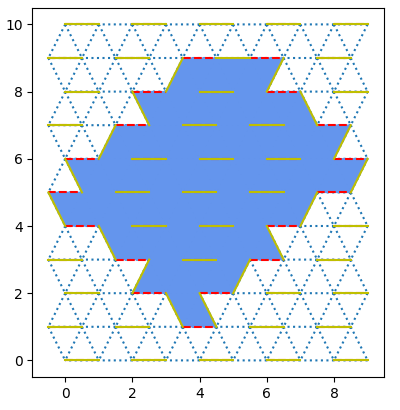} 
    \label{fig:droplet} }
    \caption{
    \subref{fig:reference} Reference configuration $\mathcal D_0$ (red dotted lines); 
    \subref{fig:droplet} example of a droplet; interior the composition cycle $(\mathcal D_0, \mathcal D)$ is in blue, here $\mathcal D$ is an optimal configuration (green solid lines). } 
    \label{fig:example}
\end{figure}

In the present work, we are using two dimer configurations: one is constant, from now on called reference configuration (see Figure~\ref{fig:reference}), and the other is the optimal configuration under chosen conditions (energies of the dimers, temperature, etc.). The composition of these configurations provides closed contours (see Figure~\ref{fig:droplet}).

Four types of edges are considered here: horizontal coinciding with reference configuration, horizontal non-coinciding with reference configuration, right-leaning, and left-leaning. Edges have corresponding energies: $E_\text{even}$, $E_\text{odd}$, $E_\text{right}$, $E_\text{left}$.

\subsection{Sampling algorithm}

\begin{figure}[ht]
    \centering
    \includegraphics[height=0.3\textwidth]{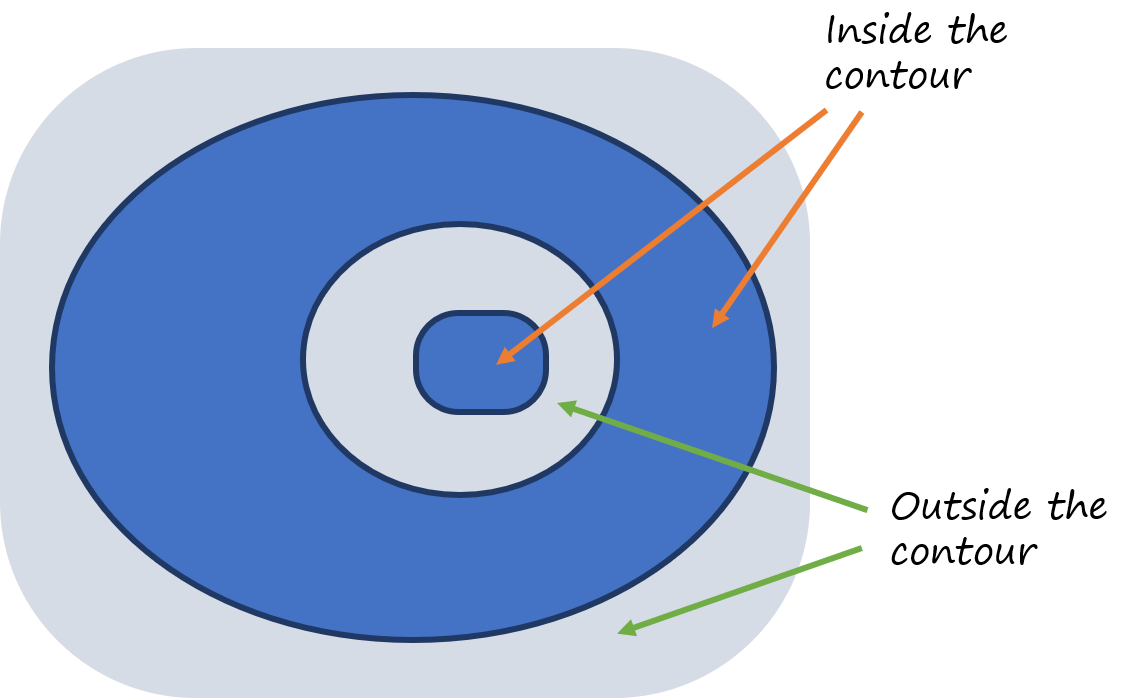}
    \caption{Notion of ``inside" the contour and ``outside".}
    \label{fig:contours}
\end{figure}

The goal of this work is to check the conjecture about the emergence of a macroscopic droplet in the dimer system. We have performed a Monte Carlo simulation within the canonical ensemble to verify this. The simulation was implemented as follows: an initial droplet of a certain area is generated, and then the dynamics of dimers is constructed through the Metropolis algorithm using the Kawasaki method~\cite{Landau-Binder-2000}, that is, in such a way that the area inside the contours remains constant.  Note that there is a natural notion of ``inside" a contour and ``outside" (see Figure~\ref{fig:contours}). As a result of such dynamics, the initial macroscopic droplet can either preserve the equilibrium shape, which confirms our hypothesis, or can be divided into many small droplets.

Dimers can not be turned by one because of perfect matching that is every vertex belongs to a dimer, and when only one dimer is turned one of its ends (vertices) will belong to an extra dimer. Consequently, dimers should be turned, for example, in pairs. However, as previous studies have shown, so-called \textit{staggered configurations}~\cite{Moessner-Sondhi-2001} may occur causing the impossibility of choosing two dimers able to be flipped. To solve this problem flips of four dimers should be added.

\begin{figure}[h]
    \vspace{2ex}
    \centering
    \includegraphics[width=1\linewidth]{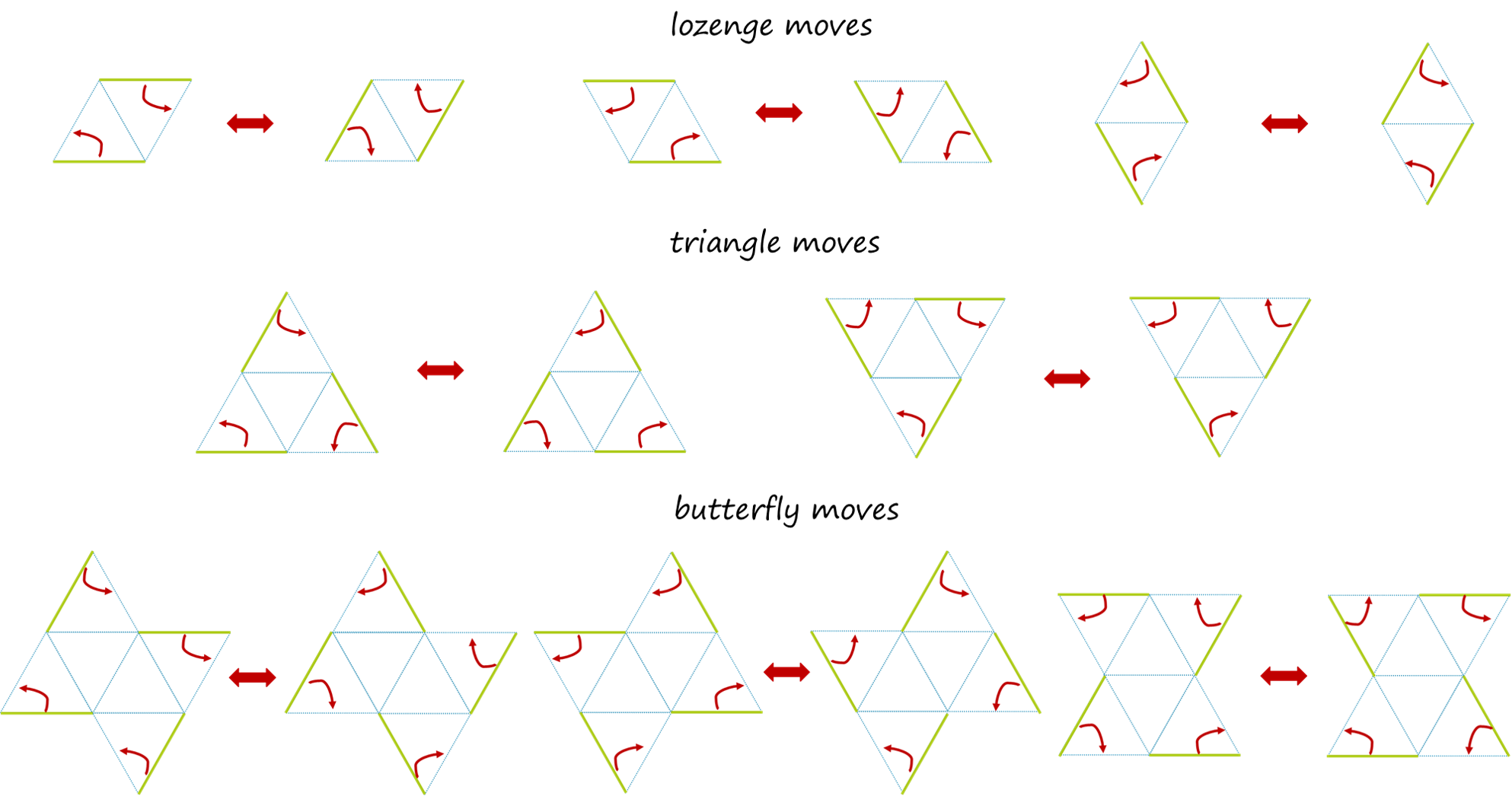}
    \caption{Types of the flippable dimers.}
    \label{fig:scheme}
\end{figure}

Kenyon and Rémila in~\cite{Kenyon-Remila-1996} demonstrated that there is a finite set of types of moves to ensure any dimer configuration: lozenge moves, triangle moves, and butterfly moves (see Figure~\ref{fig:scheme}, the reference configuration is not illustrated here). Nonetheless, Røising and Zhang proved in~\cite{Røising-Zhang-2023} that triangle moves are redundant due to the equivalency of these moves to a combination of lozenge and butterfly moves. Therefore in this work, we considered the turns of only pairs and fours of dimers whose movements do not cause a change of the initial area inside the contour. It is worth noting during the process of choosing dimers, the reference configuration should be taken into account to prevent the destruction of the existing contours. Dynamics was constructed the following way: choose randomly a pair or four dimers to flip, then, if necessary, choose, again, randomly another flippable group of dimers to compensate for changing the total area inside the contours.

\section{Simulation results}

\subsection{Observation of droplets}

\begin{figure}[h]
    \vspace{2ex}
    \centering
    \subfigure[]{
    \includegraphics[trim={3cm 0 3cm 0},clip,width=0.4\textwidth]{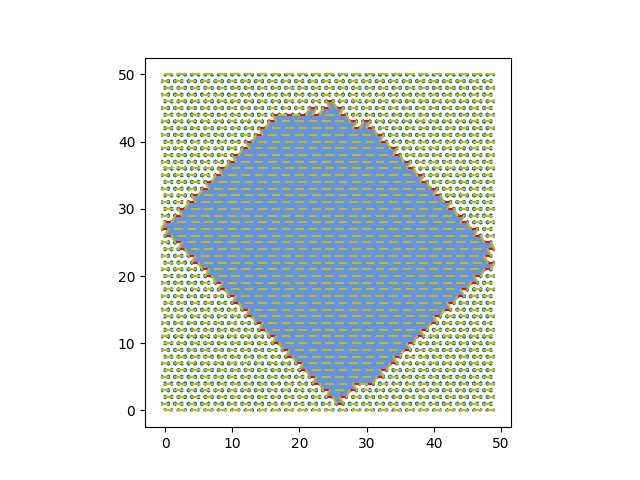} 
    \label{fig:3-1-a} } 
    \hspace{4ex}
    \subfigure[]{
    \includegraphics[trim={3cm 0 3cm 0},clip,width=0.4\textwidth]{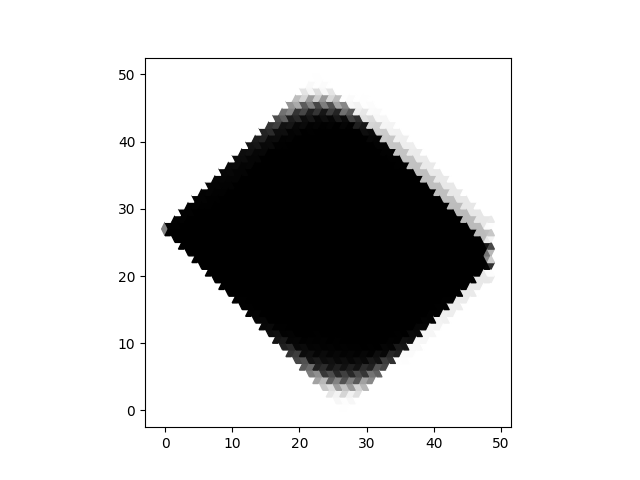} 
    \label{fig:3-1-b} }
    \caption{Area of the initial droplet is 50\% of the total area of the lattice, temperature 0.1:
    \subref{fig:3-1-a} an optimal droplet at the given temperature; 
    \subref{fig:3-1-b} the probability of being a triangle inside the contour. } 
    \label{fig:results1}
\end{figure}

We have established that a macroscopic droplet is produced with the following energies $E_\text{even} = 0$, $E_\text{odd} = 2$, $E_\text{right} = E_\text{left} = 1$. In Figure~\ref{fig:results1} an example of droplets for grid size $50\times50$ is presented. Here, $10^7$ iterations of the Metropolis algorithm were chosen for thermalisation and $10^8$ for measurements. Additionally, there is shown the probability of being inside the contour which was calculated in such a way:

\begin{equation}
    P = \frac{\text{number of steps when triangle was inside the contour}}{\text{total number of steps}}
\end{equation}
As we can see on average the shape of a droplet can be approximated by a rectangle with smoothed corners.

\subsection{Phase transition}

\begin{figure}[h]
    \vspace{2ex}
    \centering
    \includegraphics[width=1\linewidth]{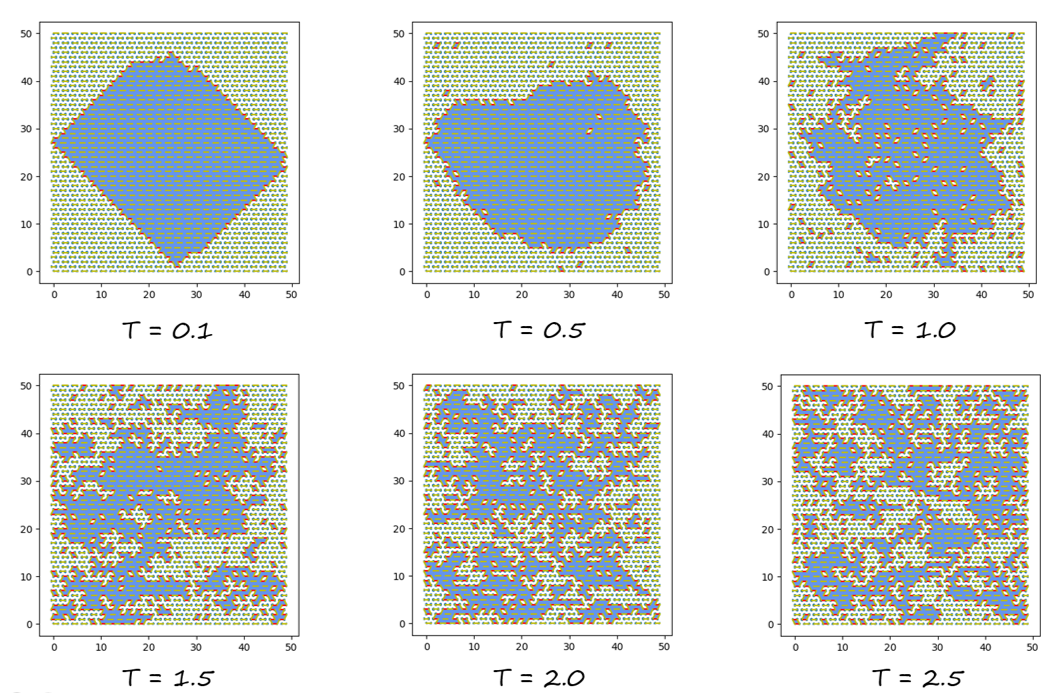}
    \caption{Phase transition in the dimer model in the triangular lattice (area of the initial droplet is 50\% of the total area of the lattice).}
    \label{fig:results2}
\end{figure}

Further, we have studied the behaviour of the initial droplet with an increase in temperature. Figure~\ref{fig:results2} shows modifications in the state of the system with an initial droplet area of 50\% of the lattice area when temperature changes. At low temperatures, one microscopic droplet is observed. Then, with increasing temperature, the droplet loses its equilibrium shape, and the droplets of the opposite phase are immersed in the initial droplet. Then the macroscopic droplet begins to break down. As a result, there are a lot of small droplets of irregular shape.

\section{Conclusions}

The main results of this work can be summarised as follows:
\begin{enumerate}[-] 
    \item it was determined that the dimer system at a low temperature, with a fixed total area within superposition polygons and with free boundary conditions tends to form a macroscopic droplet with a shape close to rectangular on average. Thus, in the dimer model, a phenomenon similar to the Dobrushin-Kotecký-Shlosman droplets in the Ising model was detected;
    \item the presence of a phase transition was also established, which is accompanied by the deformation of the equilibrium form of the macroscopic droplet with its subsequent decomposition which starts at a temperature above 0.5.
\end{enumerate}
The next step in studying this phenomenon is constructing a variational problem to determine the exact formula for the limiting shape of a droplet.

\section*{Acknowledgments}
We are grateful to Nicolai Reshetikhin and Senya Shlosman for productive discussions.

The study was carried out with the financial support of the Ministry of Science and Higher Education of the Russian Federation in the framework of a scientific project under agreement No. 075-15-2024-631.

\newpage
\addcontentsline{toc}{section}{References}
\bibliographystyle{ieeetr}
\bibliography{refs} 

\begin{thebibliography}{10}

\bibitem{Roberts-1935}
J.~K. Roberts, ``{The adsorption of hydrogen on tungsten},'' {\em Proceedings of the Royal Society of London. Series A - Mathematical and Physical Sciences}, vol.~152, pp.~445 -- 463, 1934.

\bibitem{Fowler-Rushbrooke-1937}
R.~H. Fowler and G.~S. Rushbrooke, ``{An attempt to extend the statistical theory of perfect solutions},'' {\em Transactions of The Faraday Society}, vol.~33, pp.~1272--1294, 1937.

\bibitem{Kasteleyn-1963}
P.~W. Kasteleyn, ``{Dimer Statistics and Phase Transitions},'' {\em Journal of Mathematical Physics}, vol.~4, pp.~287--293, 1963.

\bibitem{Kenyon-Okounkov-2006}
R.~W. Kenyon and A.~Okounkov, ``{Planar dimers and Harnack curves},'' {\em Duke Mathematical Journal}, vol.~131, pp.~499--524, 2003.

\bibitem{Kenyon-Okounkov-Sheffield-2003}
R.~W. Kenyon, A.~Okounkov, and S.~Sheffield, ``{Dimers and amoebae},'' {\em Annals of Mathematics}, vol.~163, pp.~1019--1056, 2003.

\bibitem{Cimasoni-Reshetikhin-2007}
D.~Cimasoni and N.~Reshetikhin, ``{Dimers on Surface Graphs and Spin Structures}. {I},'' {\em Communications in Mathematical Physics}, vol.~275, pp.~187--208, 2006.

\bibitem{Cimasoni-Reshetikhin-2008}
D.~Cimasoni and N.~Reshetikhin, ``{Dimers on Surface Graphs and Spin Structures}. {II},'' {\em Communications in Mathematical Physics}, vol.~281, pp.~445--468, 2007.

\bibitem{Fan-Wu-1970}
C.~Fan and F.~Y. Wu, ``{General Lattice Model of Phase Transitions},'' {\em Physical Review B}, vol.~2, pp.~723--733, 1970.

\bibitem{Fisher-1966}
M.~E. Fisher, ``{On the Dimer Solution of Planar Ising Models},'' {\em Journal of Mathematical Physics}, vol.~7, pp.~1776--1781, 1966.

\bibitem{Temperley-Fisher-1961}
H.~N.~V. Temperley and M.~E. Fisher, ``{Dimer problem in statistical mechanics -- an exact result},'' {\em Philosophical Magazine}, vol.~6, pp.~1061--1063, 1961.

\bibitem{McCoy-Wu-1973}
B.~M. McCoy and T.~T. Wu, {\em The Two-Dimensional Ising Model}.
\newblock Cambridge, MA and London, England: Harvard University Press, 1973.

\bibitem{Dobrushin-Kotecky-Shlosman-1992}
R.~L. Dobrushin, R.~Koteck{\'y}, and S.~B. Shlosman, {\em {Wulff Construction: A Global Shape from Local Interaction}}.
\newblock American Mathematical Society, 1992.

\bibitem{Dobrushin-1973}
R.~L. Dobrushin, ``{Gibbs State Describing Coexistence of Phases for a Three-Dimensional Ising Model},'' {\em Theory of Probability and Its Applications}, vol.~17, pp.~582--600, 1973.

\bibitem{Fröhlich-Pfister-1987}
J.~Fr{\"o}hlich and C.~E. Pfister, ``{The Wetting and Layering Transitions in the Half-Infinite Ising Model},'' {\em EPL}, vol.~3, pp.~845--852, 1987.

\bibitem{Ioffe-Ott-Shlosman-Velenik-2022}
D.~Ioffe, S.~Ott, S.~B. Shlosman, and Y.~Velenik, ``{Critical prewetting in the 2d Ising model},'' {\em The Annals of Probability}, 2020.

\bibitem{Pfister-Velenik-1997}
C.-E. Pfister and Y.~A. Velenik, ``{Mathematical theory of the wetting phenomenon in the 2D Ising model},'' {\em Helvetica Physica Acta}, vol.~69, pp.~949--973, 1997.

\bibitem{Wulff-1901}
G.~Wulff, ``{XXV}. {Zur Frage der Geschwindigkeit des Wachsthums und der Aufl{\"o}sung der Krystallfl{\"a}chen},'' {\em Zeitschrift f{\"u}r Kristallographie -- Crystalline Materials}, vol.~34, pp.~449 -- 530, 1901.

\bibitem{Abraham-Gallavotti-Martin-Lof-1971}
D.~B. Abraham, G.~Gallavotti, and A.~Martin-Lof, ``{Surface tension in the Ising model},'' {\em Lettere al Nuovo Cimento (1971-1985)}, vol.~2, pp.~143--146, 1971.

\bibitem{Abraham-Reed-1977}
D.~B. Abraham and P.~Reed, ``{Diagonal interface in the two-dimensional Ising ferromagnet},'' {\em Journal of Physics A}, vol.~10, 1977.

\bibitem{Herring-1951}
C.~Herring, ``{Some Theorems on the Free Energies of Crystal Surfaces},'' {\em Physical Review}, vol.~82, pp.~87--93, 1951.

\bibitem{Ioffe-Schonmann-1998}
D.~Ioffe and R.~H. Schonmann, ``{Dobrushin–Koteck{\'y}–Shlosman Theorem up to the Critical Temperature},'' {\em Communications in Mathematical Physics}, vol.~199, pp.~117--167, 1998.

\bibitem{Ioffe-1995}
D.~Ioffe, ``{Exact large deviation bounds up to Tc for the Ising model in two dimensions},'' {\em Probability Theory and Related Fields}, vol.~102, pp.~313--330, 1995.

\bibitem{Greenberg-Ioffe-2005}
L.~Greenberg and D.~Ioffe, ``{On an invariance principle for phase separation lines},'' {\em Annales De L Institut Henri Poincare-probabilites Et Statistiques}, 2005.

\bibitem{Campanino-Ioffe-Velenik-1995}
M.~Campanino, D.~Ioffe, and Y.~Velenik, ``{Ornstein-Zernike theory for finite range Ising models above Tc},'' {\em Probability Theory and Related Fields}, vol.~125, pp.~305--349, 2001.

\bibitem{Ioffe-1994}
D.~Ioffe, ``{Large deviations for the 2D ising model: A lower bound without cluster expansions},'' {\em Journal of Statistical Physics}, vol.~74, pp.~411--432, 1994.

\bibitem{Bricmont-Lebowitz-1987}
J.~Bricmont and J.~L. Lebowitz, ``{Wetting in Potts and Blume-Capel models},'' {\em Journal of Statistical Physics}, vol.~46, pp.~1015--1029, 1987.

\bibitem{Coninck-Messager-Miracle-Sole-Ruiz-1988}
J.~de~Coninck, A.~Messager, S.~Miracle-Sole, and J.~Ruiz, ``{A study of perfect wetting for Potts and Blume-Capel models with correlation inequalities},'' {\em Journal of Statistical Physics}, vol.~52, pp.~45--60, 1988.

\bibitem{Hryniv-Kotecky-2002}
O.~Hryniv and R.~Koteck{\'y}, ``{Surface Tension and the Ornstein–Zernike Behaviour for the 2D Blume–Capel Model},'' {\em Journal of Statistical Physics}, vol.~106, pp.~431--476, 2002.

\bibitem{Messager-Miracle-Sole-Ruiz-Shlosman-1991}
A.~Messager, S.~Miracle-Sole, J.~Ruiz, and S.~B. Shlosman, ``{Interfaces in the Potts model II: Antonov's rule and rigidity of the order disorder interface},'' {\em Communications in Mathematical Physics}, vol.~140, pp.~275--290, 1991.

\bibitem{Gartner-2014}
M.~Gartner, ``{Classical and Quantum Dimer Models},'' 2014.
\newblock \url{https://physics.ucsd.edu/~mcgreevy/s14/final-papers/2014S-239a-Gartner-Mike.pdf}.

\bibitem{Landau-Binder-2000}
D.~P. Landau and K.~Binder, {\em {A Guide to Monte Carlo Simulations in Statistical Physics}}.
\newblock Cambridge university press, 2021.

\bibitem{Fendley-Moessner-Sondhi-2002}
P.~Fendley, R.~Moessner, and S.~L. Sondhi, ``{Classical dimers on the triangular lattice},'' {\em Physical Review B}, vol.~66, p.~214513, 2002.

\bibitem{Moessner-Sondhi-2001}
R.~Moessner and S.~L. Sondhi, ``{Resonating valence bond phase in the triangular lattice quantum dimer model.},'' {\em Physical review letters}, vol.~86 9, pp.~1881--4, 2000.

\bibitem{Kenyon-Remila-1996}
C.~Mathieu and {\'E}.~R{\'e}mila, ``{Perfect matchings in the triangular lattice},'' {\em Discret. Math.}, vol.~152, pp.~191--210, 1996.

\bibitem{Røising-Zhang-2023}
H.~S. R{\o}ising and Z.~Zhang, ``Ergodic archimedean dimers,'' {\em SciPost Physics Core}, vol.~6, no.~3, p.~054, 2023.

\end{thebibliography}

\end{document}